\documentclass[aps,prl,amstext,amsmath,superscriptaddress,twocolumn]{revtex4-1}

\usepackage{graphicx}
\usepackage{color}
\usepackage{xspace}

\newcommand{\fref}[1]{Fig.~\ref{#1}}

\newcommand{\bak}{Ba$_{1-x}$K$_{x}$Fe$_2$As$_2$\xspace}
\newcommand{\baco}{Ba(Fe$_{1-x}$Co$_{x}$)$_2$As$_2$\xspace}

\newcommand{\spm}{$s_{\pm}$\xspace}

\newcommand{\tc}{$T_c$\xspace}

\newcommand{\tsm}{$T_{sm}$\xspace}

\begin{document}

\title{Interplay between superconductivity and itinerant magnetism in underdoped Ba$_{1-x}$K$_x$Fe$_2$As$_2$ ($x=$ 0.2) probed by the response to controlled point-like disorder}

\author{Ruslan Prozorov}
\email[Corresponding author: ]{prozorov@ameslab.gov}
\affiliation{Ames Laboratory, Ames, IA 50011, USA}
\affiliation{Department of Physics \& Astronomy, Iowa State University, Ames, IA 50011, USA}

\author{Marcin Ko\'nczykowski}
\affiliation{Laboratoire des Solides Irradi\'es, CEA/DRF/lRAMIS, {\'E}cole Polytechnique, CNRS, Institut Polytechnique de Paris, F-91128 Palaiseau, France}

\author{Makariy~A.~Tanatar}
\affiliation{Ames Laboratory, Ames, IA 50011, USA}
\affiliation{Department of Physics \& Astronomy, Iowa State University, Ames, IA 50011, USA}

\author{Hai-Hu Wen}
\affiliation{Center for Superconducting Physics and Materials,
National Laboratory of Solid State Microstructures $\&$ Department of Physics,\\Nanjing University, Nanjing 210093, China}

\author{Rafael M. Fernandes}
\affiliation{School of Physics and Astronomy, University of Minnesota, Minneapolis,
MN 55455, USA}

\author{Paul C.~Canfield}
\affiliation{Ames Laboratory, Ames, IA 50011, USA}
\affiliation{Department of Physics \& Astronomy, Iowa State University, Ames, IA 50011, USA}

\date{20 June 2019}

\begin{abstract}
The response of superconductors to controlled introduction of point-like disorder is an important tool to probe their microscopic electronic collective behavior. In the case of iron-based superconductors (IBS), magnetic fluctuations presumably play an important role in inducing high temperature superconductivity. In some cases, these two seemingly incompatible orders coexist microscopically. Therefore, understanding how this unique coexistence state is affected by disorder can provide important information about the microscopic mechanisms involved. In one of the most studied pnictide family, hole-doped Ba$_{1-x}$K$_x$Fe$_2$As$_2$ (BaK122), this coexistence occurs over a wide range of doping levels, 0.16~$\lesssim x \lesssim $~0.25. We used relativistic 2.5 MeV electrons to induce vacancy-interstitial (Frenkel) pairs that act as efficient point-like scattering centers. Upon increasing dose of irradiation, the superconducting transition temperature $T_c$ decreases dramatically. In the absence of nodes in the order parameter this provides a strong support for a sign-changing $s_{\pm}$ pairing. Simultaneously, in the normal state, there is a strong violation of the Matthiessen's rule and a decrease (surprisingly, at the same rate as $T_c$) of the magnetic transition temperature $T_{sm}$, which indicates the itinerant nature of the long-range magnetic order. Comparison of the hole-doped BaK122 with electron-doped Ba(Fe$_x$Co$_{1-x}$)$_2$As$_2$ (FeCo122) with similar $T_{sm}\sim$110~K, $x=$0.02, reveals significant differences in the normal states, with no apparent Matthiessen's rule violation above $T_{sm}$ on the electron-doped side. We interpret these results in terms of the distinct impact of impurity scattering on the competing itinerant antiferromagnetic and $s_{\pm}$ superconducting orders.

\end{abstract}

\maketitle

\section*{Introduction}

The use of controlled disorder is a powerful phase-sensitive way to study the nature of the superconducting state without affecting the chemical composition \cite{Anderson1959JPCS,Abrikosov1960,Hirschfeld1993PRB,KimMuzikar1994PRB,Balatsky2006RMP,Prozorov2011RPP,WangHirschfeldMishra2013PRB,Prozorov2014PRX,Mizukami2014NatureComm,Shibauchi2017}. According to Anderson's theorem \cite{Anderson1959JPCS}, conventional isotropic $s-$wave superconductors are not affected by the scalar potential (i.e. non spin-flip) scattering, but are sensitive to spin-flip scattering due to magnetic impurities (for recent theoretical results on the impact of impurities on $T_c$, see for example Refs.\cite{Kang2016,Gastiasoro2017,Kivelson2018}). In single-band high superconducting transition temperature (high-$T_c$) cuprates, both magnetic and non-magnetic impurities cause a rapid suppression of $T_c$,  consistent with the nodal $d-$wave pairing~\cite{Xiao1990PRB}. In multi-band iron-based superconductors (IBS), a sign-changing order parameter between the electron-like and hole-like Fermi sheets, \spm, is the most plausible pairing state ~\cite{MazinSPM2008,Mazin2009PhysicaC,Hirschfeld2011ROPP,Chubukov2012ARCMP,Chubukov-Hirschfeld-PT_2015,Hirschfeld2016}. Although its response to non-magnetic scattering depends sensitively on the multi-band  structure of the pairing interaction, on the chemical potential, and on the gap anisotropy, it is generally expected that inttaband scattering is much less efficient in causing pair-breaking than interband scattering  \cite{Mishra2009PRB,Onari2009PRL,Onari2010,Kontani2010PRL,Efremov2011,Chubukov2012ARCMP,FernandesChubukow2012PRB,WangHirschfeldMishra2013PRB,Xen2016,Trevisan2018}. Additionally, the orbital content of the bands can also affect the suppression of $T_c$ \cite{Onari2009PRL,Kontani2010PRL,Kreisel2015,BaekBuchner2015NatMat_NMR,Chubukov2016,Hoyer2015}.
We note that the multi-band character of the superconducting state alone is not sufficient to have $T_c$ suppression \cite{Golubov97}. For instance, in the known two-gap $s_{++}$ superconductor, MgB$_2$, where the gap does not change sign, electron irradiation resulted only in a small change due to gap magnitude difference between two bands \cite{KleinPRL2010MgB2}.

While the effect of scattering induced by various means from chemical substitution to irradiation with various particles on $T_c$ has been studied in many IBS, there is limited experimental information on the effects of point-like disorder simultaneously on superconducting and magnetic transitions in the regime where superconductivity and antiferromagnetism coexist. The expected physics, however, is very intriguing. Assuming an itinerant nature for long-range magnetic order (LRMO), it has been shown that $T_c$ may actually increase upon the introduction of disorder due to the stronger effect on magnetism quantified via the suppression of the magnetic transition temperature, $T_{sm}$. (Here ``$sm$" is used to indicate simultaneous structural and magnetic transitions in underdoped BaK122) \cite{FernandesChubukow2012PRB}. However, this is not a universal trend, as it depends on the relative ratio of the magnetic and superconducting state energies and on the relative strength of the intraband and interband scattering rates.

Irradiation of relatively thin crystals ($\sim$20 $\mu$m in our case) with 2.5 MeV relativistic electrons is known to produce vacancy - interstitial Frenkel pairs, which act as efficient point-like scattering centers \cite{Damask1963,THOMPSON1969}. In the high-\(T_{c}\) cuprates these defects are known to be strong unitary scatterers causing significant suppression of \(T_{c}\) \cite{Rullier-Albenque2003PRL}. There is a growing number of studies of the effects of electron irradiation not only on $T_c$ (see \cite{Prozorov2014PRX} and references therein), but on other properties, such as vortex pinning and creep \cite{VanDerBeek2013JPCS_M2S} and London penetration depth \cite{Mizukami2014NatureComm,Strehlow2014,Cho2014PRB,Cho2016BaK,Teknowijoyo2016}. In a previous study of electron irradiated \bak~ we focused on the evolution of the superconducting gap structure with the potassium concentration and found noticeable changes in the behavior, such as increasing gap anisotropy \cite{Cho2014PRB,Cho2016BaK,SUSTreview}.

In this paper, we focus on the effects of electron irradiation simultaneously on \tc, \tsm~ and normal state resistivity in an underdoped composition of (Ba$_{1-x}$K$_x$)Fe$_2$As$_2$, $x=0.2$ \cite{phaseD}, in which LRMO coexists with superconductivity.  In the normal state, we find  strong violation of the Matthiessen's rule below \tsm, which is expected {due to a change of the band structure and thus effective carrier density}  in the magnetically ordered state, and above \tsm~ in the {broad} 
temperature range {,} which is unexpected at least in the simple picture. Moreover, this behavior is in a stark contrast with the electron-doped \baco~ ($x=0.02$) with similar \tsm~ in which the Matthiessen's rule is expectedly violated below \tsm~ but obeyed above \tsm. At a first sight this could be understood that in this case additional disorder is not so effective, because despite notably lower substitution level, $x$, required to suppress magnetism, doping into Fe-As planes introduces much higher scattering rates as evidenced by notably higher residual resistivity values. This argument, however, does not hold, because (1) the magnetic transition temperature, $T_{sm}$~, changes by a similar {amount} as in BaK122 and (2) the same compliance with the Matthiessen's rule above $T_{sm}$ is observed in isovalently substituted Ba(Fe$_{1-x}$Ru$_{x}$)$_2$As$_2$ \cite{Blomberg2018} and very clean BaFe$_2$(As$_{1-x}$P$_{x}$)$_2$ \cite{Shibauchi2017}. Therefore, the difference is likely in the electronic structure of BaK122 and specifics of its inter- and intra-band interactions and scattering channels \cite{FernandesChubukow2012PRB}.

\begin{figure}[tb]
\includegraphics[width=9cm]{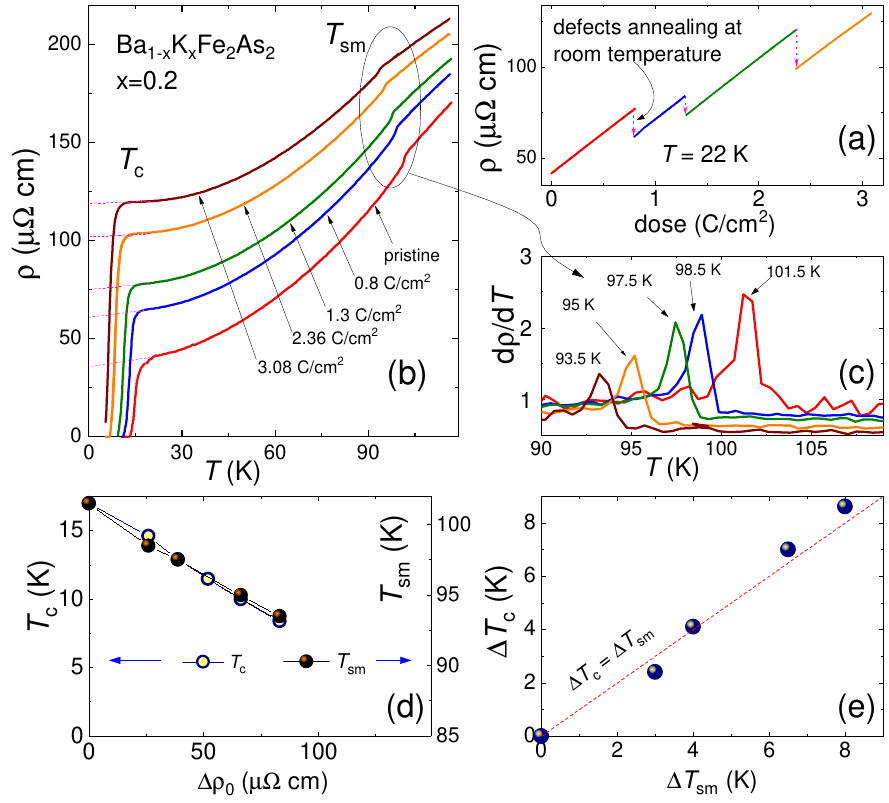}
\caption{{\bf Effect of repeated electron irradiation on resisitivity.} (a) Dose dependence of  electrical resistivity {of sample $A$} at 22~K measured {\it in-situ} during electron irradiation. The resistivity increases linearly during single irradiation run, steps in the curve result from partial defect annealing on warming sample to room temperature between runs for characterization. (b) Temperature-dependent resistivity, $\rho (T)$, measured after subsequent irradiation runs and room-temperature annealing. (c) Temperature-dependent resistivity derivative, $d \rho /dT$, revealing a sharp peak at \tsm. Peak position was used as a criterion for \tsm~ determination. (d) Variation of the superconducting transition temperature, \tc~ (left axis, determined at onset cross-point) and magnetic/structural transition, \tsm~ (right axis, determined from resistivity derivative peak) offset vertically to match each other (see text). (e) Variation of \tc~ plotted as a function of the variation of \tsm.}
\label{fig1}
\end{figure}

\section*{Results and Discussion}

The panel (b) in Fig.~\ref{fig1} shows the evolution of the temperature dependent resistivity of under-doped Ba$_{1-x}$K$_x$Fe$_2$As$_2$, $x$=0.20, with increase of irradiation dose/disorder. We zoom on the low-temperature range revealing features in $\rho(T)$ curves at the structural/magnetic,  \tsm, and superconducting,  \tc, transitions. The resistivity of the samples at temperatures just above \tc~ follows well a $\rho(0)+AT^2$ dependence, similar to previous reports \cite{Xigang}, allowing easy extrapolation of $\rho(0)$ and tracking its evolution with disorder. The residual resistivity of the pristine samples was about 40 $\mu \Omega$cm, and residual resistivity ratio $\rho(300K)/\rho(0) >$7. Residual resistivity increased up to approximately 120 $\mu \Omega$cm at the highest dose of 6 C/cm$^2$. On structural/magnetic ordering, resistivity of the sample shows small down-turn on cooling, due to a loss of spin-disorder scattering. The structural transition temperature was determined using temperature dependent resistivity derivative and peak position as a criterion, as shown in Fig.~\ref{fig1} (c). The \tsm~ is monotonically suppressed with increase of sample residual resistivity  as shown in Fig.~\ref{fig1} panel (d), right scale. The superconducting transition temperature was determined at crossing points of linear extrapolations of the sharp resistivity drop at the transition and smooth $T^2$ extrapolations of the curves in the normal state. The irradiation does not change the sharpness of the resistive transition, so the use of the alternative criterion (midpoint or zero resistance cross-point) for \tc~ determination does not alter any of our conclusions. The \tc~  shows monotonic decease with irradiation from above 16 to 9~K, Fig.~\ref{fig1} (d). Interestingly, the decrease of both temperatures in absolute numbers is almost the same and the two are linearly proportional to each other, see Fig.~\ref{fig1} panel (e).

\begin{figure}[tb]
\includegraphics[width=8.5cm]{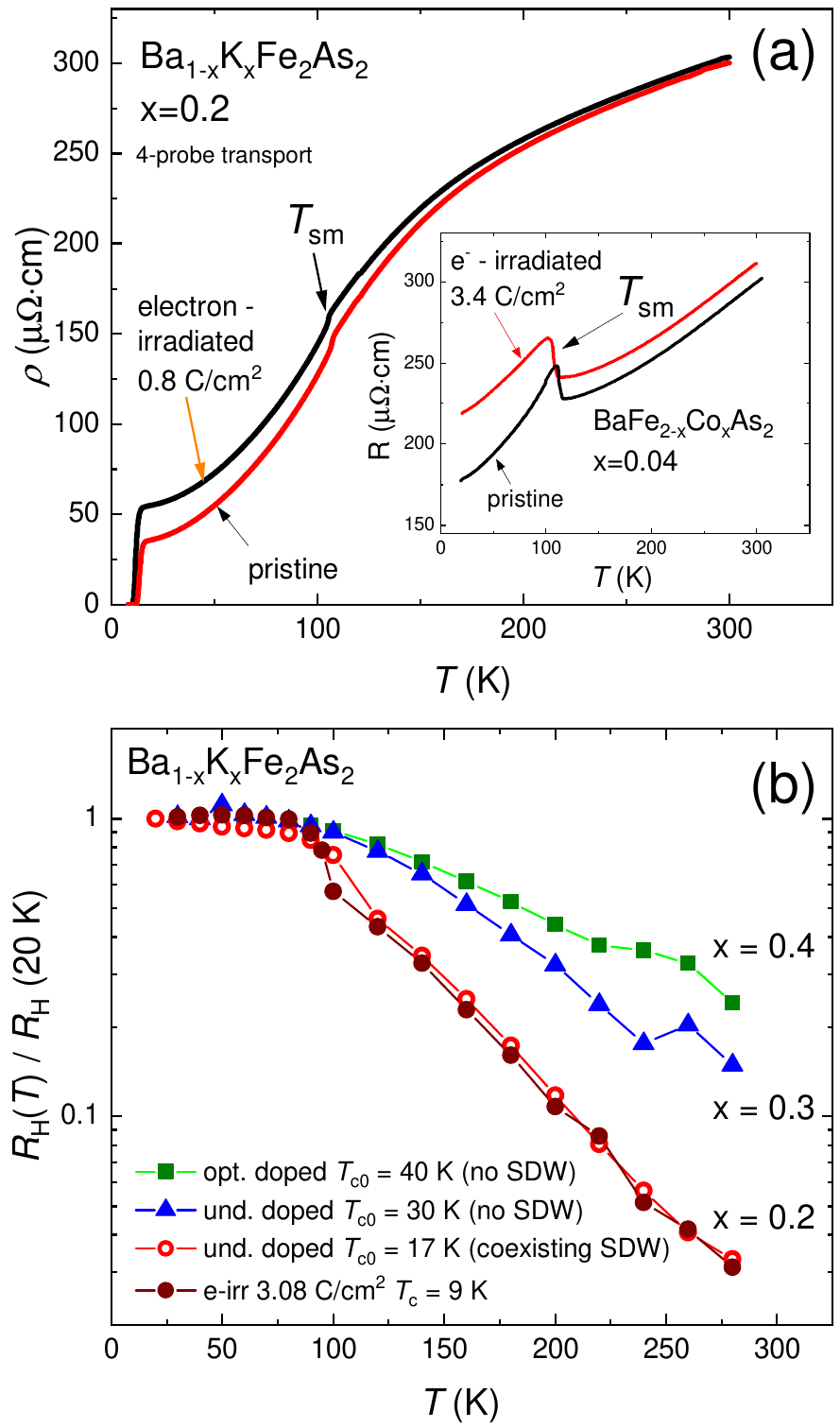}
\caption{{\bf Electric resistivity and Hall effect:} (a) Comparison of the effect of electron irradiation on the normal-state resistivity of hole (sample $B$ of BaK122, main panel) and electron (sample $C$ of BaCo122, inset) - doped Ba122 measured using 4-probe method. The peak-like character of the magnetic transition indicates much more background scattering in the electron-doped compound. (b) Normalized Hall coefficient obtained from van der Pauw measurements as function of temperature before (open circles) and after (filled circles) electron irradiation of sample $A$ for $x=$0.2 showing no change. This is compared to the data for $x=$0.3 (triangles) and 0.4 (squares) showing clear doping dependence. }
\label{fig2}
\end{figure}

The top panel (a) of Fig.~\ref{fig2} shows 4-probe resistivity measured in the sample $B$ of Ba$_{1-x}$K$_x$Fe$_2$As$_2$, $x$=0.20, in pristine state and after electron irradiation with the dose of 0.8 C/cm$^2$. Even at this relatively small dose, a clearly non-parallel shift of the curves indicates significant violation of the  Matthiessen's rule in the paramagnetic state.
In Drude model, the electrical conductivity of metal $\sigma= en\frac{\tau}{m^*}$, here $n$ is the carrier density, $\tau$ is the scattering time, $e$ is electron charge and $m^*$ is the effective bare mass of the carriers. The Matthiessen's rule states that the total scattering rate $\tau^{-1}$ is the sum of the scattering rates of different contributions, elastic and inelastic, $\tau^{-1}= \tau_0^{-1}+\tau_i^{-1}$, which leads to a parallel shift of $\rho(T)$ curve with the increase of residual $\rho(0)$ with disorder. The Matthiessen's rule can be violated by Fermi surface reconstruction, which takes place below $T_{sm}$, or even above $T_{sm}$ as a consequence of the anisotropic character of the magnetic fluctuations \cite{Stephen1977}.  Close to room temperature, the Matthiessen's rule is valid, with the violation being closely linked with a crossover feature in the temperature-dependent resistivity at around 200~K. For comparison in inset in panel (a), Fig.~\ref{fig2}, we show temperature-dependent resistivity of slightly electron doped Ba(Fe$_{1-x}$Co$_x$)$_2$As$_2$, $x=$0.02, in pristine state and after 3.4 C/cm$^2$ electron irradiation. Notably, electron irradiation leads to a comparable increase of the room temperature resistivity, $\rho$(300 K), in both electron and hole-doped compositions, around 3-4 $\mu \Omega$cm per C/cm$^2$, respectively. However, the increase remains practically constant above \tsm~ in electron-doped composition, similar to the behavior of hole-doped composition above the crossover feature and to phosphorus-substituted samples \cite{Shibauchi2017}. Also, the resistivity change at \tsm upon cooling is quite different between hole and electron doped compounds, showing only a slight downturn in the former, but a pronounced jump in the latter.

To understand the difference in behavior, it is important to note that orbitals of iron and arsenic in FeAs layer are contributing the most to the density of states at the Fermi level in BaFe$_2$As$_2$ based materials. Therefore disorder introduced by random positions of substitutional Co atoms in the FeAs layer, affects electron scattering significantly stronger than substitutional disorder of K on Ba cite. This can be directly seen in notably lower residual resistivity in Ba$_{1-x}$K$_x$Fe$_2$As$_2$, $x$=0.20, $\rho(0)\sim$40 $\mu\Omega$cm than in Ba(Fe$_{1-x}$Co$_x$)$_2$As$_2$, $x=$0.02, $\rho(0)\sim$170 $\mu\Omega$cm, despite five times smaller level of substitution in the latter case. Because of this high level of substitutional disorder in Co-doped material, additional disorder introduced by electron irradiation plays relatively smaller role than in K-doped compound.
This different level of background disorder leads to different resistivity behavior on passing \tsm~ in pristine samples. Loss of the carrier density below \tsm~ due to partial gap opening in conditions when carrier mean free path is controlled by disorder and is essentially temperature-independent, gives resistivity increase in disordered Co-doped material. Same loss is compensated by notable increase of mean free path due to the loss of spin-disorder scattering in hole-doped compositions. These considerations were directly illustrated recently in irradiation study on BaFe$_2$As$_2$ with P substitution \cite{Shibauchi2017}. Note, however, that these considerations do not explain the violation of the Matthiessen rule above $T_{sm}$ in hole-doped as opposed to its validity in electron-doped compositions. The difference is not related to the level of substitutional disorder in two cases, since both absolute increase of resistivity above $T_{sm}$ and suppression rate of $T_{sm}$ with disorder are very similar on both sides.

The lower panel (b) of Fig.~\ref{fig2} shows temperature dependent Hall coefficient, $R_H$, obtained using van der Pauw technique in the sample $A$ of Ba$_{1-x}$K$_x$Fe$_2$As$_2$, $x=$0.20 before and after irradiation. For reference we show data in other hole-doped samples, $x=$0.3 and $x=$0.4, in all cases normalizing data at 20~K, the lowest temperature of our Hall effect measurements. (For x=0.4 and x=0.3, the curves were extrapolated to $T=$20~K). Irradiation does not change either magnitude  or temperature dependence of the Hall effect in sample with $x=$0.20, despite three-fold variation of sample the resistivity. On the other hand doping clearly changes magnitude and temperature dependence of the Hall effect. These observations clearly show that defects introduced by irradiation are not doping the system. It should be noted that the independence of Hall coefficient on the residual resistivity is possible only if all types of carriers change their mobility by the same factor, - not so easy condition to meet in multi-band systems \cite{Blomberg2018}.

\begin{figure}[tb]
\includegraphics[width=8.5cm]{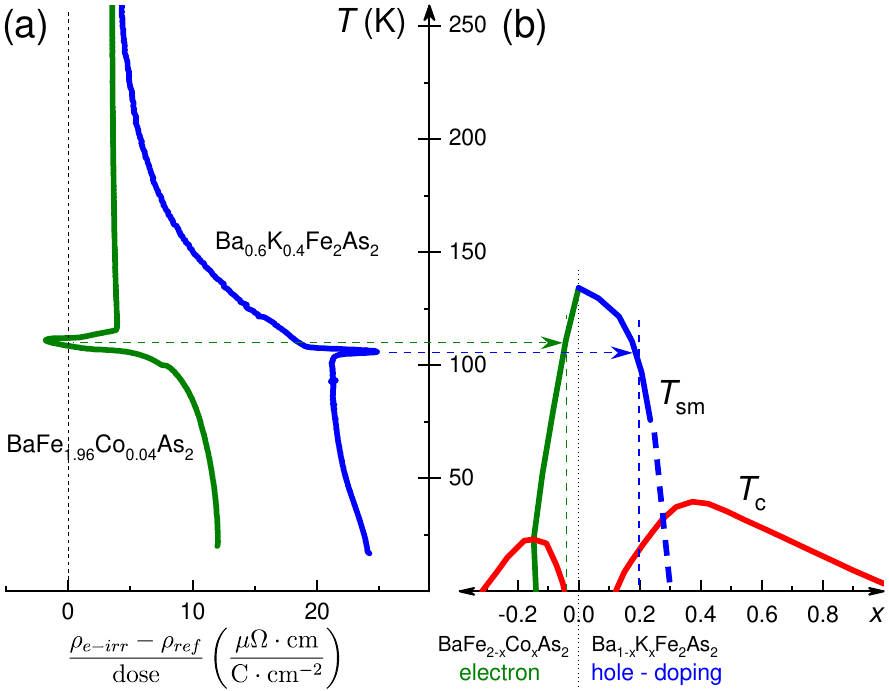}
\caption{{\bf Matthiessen's rule:} (a) Temperature dependence of the resistivity difference before and after irradiation normalized by the irradiation dose showing gross violation of the Matthiessen's rule in BaK122 extending to temperatures well above \tsm, compared to practically a constant shift in BaCo122 in this regime. (b) Summary phase diagram of electron (BaCo122) and hole-doped (BaK122) indicating the concentrations studied in panel (a).}
\label{fig3}
\end{figure}

Figure~\ref{fig3} highlights the difference between BaCo122 and BaK122 with respect to the Matthiessen's rule. Figure~\ref{fig3}(a) shows temperature dependent difference of resistivities,  before $(\rho_{ref})$ and after $(\rho_{e-irr})$ electron irradiation, normalised by the irradiation dose, $\left(\rho_{e-irr}(T)-\rho_{ref}(T)\right)/\mathrm{dose}$. Such normalization is justified by the linear scaling of the induced additional resistivity with the dose, \fref{fig1}(a), as well as transition temperatures, \fref{fig1}(d),(e). Figure~\ref{fig3}(b) shows $T-x$ phase diagram of electron (BaCo122) and hole-doped (BaK122) using the same temperature axis as in panel (a) and indicating the compositions shown in panel (a). The resulting comparison shows highly contrasting behavior, revealing a good validity of the Matthiessen's rule in BaCo122 in the paramagnetic state, but a significant violation in BaK122 extending to temperatures well above \tsm. This asymmetry is one of the main results of this work. Together with the observed rapid suppression of \tsm and \tc with added non-magnetic scattering, described in \fref{fig1}, these results provide important experimental insights into the nature of the interplay between itinerant magnetism and high-temperature superconductivity in iron-based superconductors.

If magnetism was due to localized spins, one would expect that disorder, as introduced in our experiment, would affect $T_{sm}$ primarily via the effect of random dilution \cite{Stephen1977}. If magnetism however arises from a Fermi surface instability, the change in the lifetime of the electronic states will affect $T_{sm}$. Indeed, Ref. \cite{FernandesChubukow2012PRB}, studying a simplified two-band model for the interplay between superconductivity and magnetism, showed that both intraband and interband impurity scattering suppress $T_{sm}$. This is to be contrasted with the case of $s_{\pm}$ superconductivity, in which $T_c$ is only affected by interband impurity scattering. Because long-range magnetic order competes with superconductivity, depending on how strong this competition is, it is possible that the net effect of disorder is to increase $T_c$ in the coexistence region. This seems to be the case in P-doped Ba122 \cite{Shibauchi2017}, but not in BaK122, were we find $T_c$ to also be suppressed. One possible explanation for this difference would be that the competition between superconductivity and magnetism is not as strong in K-doped systems as in P-doped systems, or that the intraband scattering is dominant over the interband scattering in the system studied here.

As for the Matthiessen's rule, a known scenario in which it is violated is when impurities are added in a system whose main scattering mechanism is strongly anisotropic in momentum space \cite{Rosch1999}. In BaK122, a natural candidate is the scattering by spin fluctuations, which in this system are strongly peaked at the finite wave-vectors $(\pi,0)$ and $(0,\pi)$. In this case, the violation of the Matthiessen's rule would imply that the resistivity of the normal state is dominated by magnetic fluctuations. Such a preponderance of magnetic fluctuations could in principle favor a higher $T_c$ state, if indeed pairing is mediated by spin fluctuations.

It should be noted, however, that the position of hot spots as well as the strength of inelastic scattering are very similar on electron and hole doped sides, while the effect of disorder (as seen in direct comparison Fig.~\ref{fig3}) is dramatically different.
Alternative explanation for the strong violation of the Matthiessen rule was suggested in a recent study of the effect of natural growth disorder on properties of BaFe$_2$As$_2$ with Ru substitution \cite{Blomberg2018}. Here it was assigned to predominant suppression of high mobility carriers with disorder.

In conclusion, irradiation with relativistic 2.5 MeV electrons leads to rapid suppression of both superconducting $T_c$ and the temperature of concomitant orthorhombic/antiferromagnetic transition $T_{sm}$.  In the absence of nodes in the superconducting order parameter, observation of rapid suppression of $T_c$ provides a strong support for a sign-changing $s_{\pm}$ pairing. Rapid suppression of $T_{sm}$, surprisingly at the same rate as $T_c$, indicates the itinerant nature of the long-range magnetic order. Comparison of the hole-doped Ba$_{1-x}$K$_x$Fe$_2$As$_2$ with electron-doped Ba(Fe$_x$Co$_{1-x}$)$_2$As$_2$ with similar $T_{sm}\sim$110~K, $x=$0.02, reveals significant differences in the normal states, with no Matthiessen's rule violation above $T_{sm}$ on the electron-doped side and strong violation on the hole-doped side. Our results provide strong evidence of the itinerant nature of the AFM phase and non-trivial influence of {non-magnetic disorder on coupled} superconductivity and magnetism in iron based superconductors.

\section*{Methods}

\subsection*{Single crystalline samples and transport measurements}
Single crystals of (Ba$_{1-x}$K$_x$)Fe$_2$As$_2$ were synthesized using high temperature FeAs flux method \cite{HaiHuWencrystals}.  {Two samples $A$ and $B$ were used. The} electrical resistivity {and Hall} effect of
{sample $A$ were} measured on $\sim 1 \times 1$ mm$^2$ sample with four contacts soldered with Sn \cite{SUST} at the corners in van der Pauw configuration  \cite{vdP}. {The electrical resistivity of sample $B$ was measured in standard 4-probe configuration.} The resistivity {of both samples} at room temperature, $\rho(300K)$, before irradiation was set to 300 $\mu \Omega$cm
\cite{yliu2014}, the value determined from measurements on big arrays of crystals, with actually measured value being within 10 \% uncertainty of geometric factor determination.

Single crystals of Ba(Fe$_{1-x}$Co$_x$)$_2$As$_2$ were grown from
FeAs/CoAs flux from a starting load of metallic Ba, FeAs, and CoAs, as described
in detail elsewhere \cite{Cogrowth}. The composition of the sample was determined using wavelength dispersive spectroscopy (WDS) version of electron probe microanalysis as $x$=0.02$\pm$0.002. {The electrical resistivity of the sample $C$} was measured in four-probe configuration. Similar to hole-doped sample, resistivity of the sample before irradiation was set as 300 $\mu \Omega$cm \cite{pseudogap}.
The samples were mounted on a thin mica plate in a a hollow {\it Kyocera} chip, so that they could be moved between irradiation chamber and resistivity and Hall effect setups in a different $^4$He cryostat without disturbing the contacts.

\subsection*{Electron irradiation}
The low-temperature 2.5 MeV electron irradiation was performed at the SIRIUS Pelletron linear accelerator operated by the \textit{Laboratoire des Solides Irradi\'{e}s} (LSI) at the \textit{Ecole Polytechnique} in Palaiseau, France \cite{SIRIUS}. The {\it Kyocera} chip was mounted inside the irradiation chamber and was cooled by a flow of liquid hydrogen to $T \approx 22$~K to remove excess heat produced by relativistic electrons upon collision with the ions. The flux of electrons amounted to about 2.7 $\mu$A of electric current through a 5 mm diameter diaphragm. This current was measured with the Faraday cage placed behind the hole in the sample stage, so that only transmitted electrons were counted. The irradiation rate was about $5 \times 10^{-6}$ C$/$(cm$^{2}\cdot $s) and large doses were accumulated during several irradiation runs. {The resistance of sample $A$} at 22~K was monitored \textit{in-situ} during irradiation, revealing linear increase with irradiation dose \cite{Prozorov2014PRX}, one segment of the broken line in Fig.~\ref{fig1}a. Periodically {the sample $A$ was} extracted from irradiation chamber and the effect of irradiation was characterized by \textit{ex-situ} measurements of electrical resistivity as function of temperature, \fref{fig1}(b), and of the Hall effect (see Fig.~\ref{fig2}b). Warming the sample to room temperature leads to partial defect annealing, as can be seen as the down-steps in the dose dependence of resistivity at 22~K at the start of the next irradiation run. This annealing is incomplete, as evidenced by gradual increase of resistivity for subsequent runs. The resistivity of the sample at room temperature remained stable for a period of at least several months, unless the sample was further warmed above room temperature. {For sample $B$ of K-doped and sample $C$ of Co-doped materials the whole dose was applied in one shot without intermediate measurements.}

\section*{Acknowledgments}

We thank I. Mazin and J. Kang for useful discussions. We also thank N. Ni for providing high quality FeCo122 samples. This work was supported by the U.S. Department of Energy (DOE), Office of Science, Basic Energy Sciences, Materials Science and Engineering Division. The research was performed at Ames Laboratory, which is operated for the U.S. DOE by Iowa State University under contract DE-AC02-07CH11358.

We acknowledge the French EMIR (R\'eseau national d’acc\'el\'erateurs pour les \'Etudes des Mat\'eriaux sous Irradiation) network for provision of irradiation beam time under EMIR Proposals \# 13-11-0484, 15-5788, 16-4368, and 18-5354, and we would like to thank B. Boizot and the whole team for assistance in using the SIRIUS facility.

The work in China was supported by National Key R\&D Program of China (grant no. 2016YFA0300401) and National Natural Science Foundation of China (grant no. 11534005). Theoretical work (R.M.F.) was supported by the U.S. Department of Energy, under Award DE-SC0012336.

\section*{References}
\bibstyle{nature}

\newpage

\section*{Figure Legends}

\begin{figure}[tbh]
\caption{{\bf Effect of repeated electron irradiation on resisitivity.} (a) Dose dependence of  electrical resistivity {of sample $A$} at 22~K measured {\it in-situ} during electron irradiation. The resistivity increases linearly during single irradiation run, steps in the curve result from partial defect annealing on warming sample to room temperature between runs for characterization. (b) Temperature-dependent resistivity, $\rho (T)$, measured after subsequent irradiation runs and room-temperature annealing. (c) Temperature-dependent resistivity derivative, $d \rho /dT$, revealing a sharp peak at \tsm. Peak position was used as a criterion for \tsm~ determination. (d) Variation of the superconducting transition temperature, \tc~ (left axis, determined at onset cross-point) and magnetic/structural transition, \tsm~ (right axis, determined from resistivity derivative peak) offset vertically to match each other (see text). (e) Variation of \tc~ plotted as a function of the variation of \tsm.}
\label{fig1}
\end{figure}

\begin{figure}[tbh]
\caption{{\bf Electric resistivity and Hall effect:} (a) Comparison of the effect of electron irradiation on the normal-state resistivity of hole (sample $B$ of BaK122, main panel) and electron (sample $C$ of BaCo122, inset) - doped Ba122 measured using 4-probe method. The peak-like character of the magnetic transition indicates much more background scattering in the electron-doped compound. (b) Normalized Hall coefficient obtained from van der Pauw measurements as function of temperature before (open circles) and after (filled circles) electron irradiation of sample $A$ for $x=$0.2 showing no change. This is compared to the data for $x=$0.3 (triangles) and 0.4 (squares) showing clear doping dependence. }
\label{fig2}
\end{figure}

\begin{figure}[tbh]
\caption{{\bf Matthiessen's rule:} (a) Temperature dependence of the resistivity difference before and after irradiation normalized by the irradiation dose showing gross violation of the Matthiessen's rule in BaK122 extending to temperatures well above \tsm, compared to practically a constant shift in BaCo122 in this regime. (b) Summary phase diagram of electron (BaCo122) and hole-doped (BaK122) indicating the concentrations studied in panel (a).}
\label{fig3}
\end{figure}


\begin{thebibliography}{99}

\bibitem{Anderson1959JPCS} Anderson, P.~W. Theory of Dirty Superconductors. {\it J. Phys. Chem. Solids} {\bf 11}, 26-30 (1959).


\bibitem{Abrikosov1960}
Abrikosov, A. A. \& Gor’kov, L. P. Theory of Superconducting Alloys with Paramagnetic Impurities, {\it Zh. Eksp. Teor. Fiz.} {\bf 39}, 1781-1796 (1960) [Contribution to the Theory of Superconducting Alloys with Paramagnetic Impurities,
{\it Sov. Phys. JETP} {\bf 12}, 1243-1258 (1961)].


\bibitem{Hirschfeld1993PRB} Hirschfeld, P. J. \& Goldenfeld, N. Effect of strong scattering on the low-temperature penetration depth of a d-wave superconductor, {\it Phys. Rev. B} {\bf 48}, 4219-4222 (R)(1993).

\bibitem{KimMuzikar1994PRB} Kim, H., Preosti, G. \& Muzikar, P. Penetration depth and impurity scattering in unconventional superconductors: T=0 results, {\it Phys. Rev. B} {\bf 49}, 3544-3547 (1994).

\bibitem{Balatsky2006RMP} Balatsky, A. V., Vekhter, I. \& Zhu, J.-X. Impurity-induced states in conventional and unconventional superconductors, {\it Rev. Mod. Phys.} {\bf 78}, 373-433 (2006).

\bibitem{Prozorov2011RPP} Prozorov, R. \& Kogan, V. G. London penetration depth in iron-based superconductors, {\it Rep. Prog. Phys.} {\bf 74}, 124505 (2011).


\bibitem{WangHirschfeldMishra2013PRB} Wang, Y., Kreisel, A., Hirschfeld, P. J. \& Mishra, V. Using controlled disorder to distinguish $s_\pm$ and $s_{++}$ gap structure in Fe-based superconductors, {\it Phys. Rev. B} {\bf 87}, 094504 (2013).

\bibitem{Prozorov2014PRX} Prozorov, R. {\it et al.} Effect of Electron Irradiation on Superconductivity in Single Crystals of Ba(Fe$_{1-x}$Ru$_x$)$_2$As$_2$ ($x$=0.24), {\it Phys. Rev. X }{\bf 4}, 041032 (2014).

\bibitem{Mizukami2014NatureComm} Mizukami, Y. {\it et al.} Disorder-induced topological change of the superconducting gap structure in iron pnictides, {\it Nat. Comm.} {\bf 5}, 5657 (2014).

\bibitem{Shibauchi2017} Mizukami, Y. {\it et al.} Impact of Disorder on the Superconducting Phase Diagram in BaFe$_2$(As$_{1-x}$P$_x$)$_2$, {\it J. Phys. Soc. Jpn.} {\bf 86}, 083706 (2017).


\bibitem{Gastiasoro2017} Gastiasoro, M. N. Andersen, B. M. Enhancing Superconductivity by Disorder, {\it Phys. Rev. B} {\bf 98}, 184510 (2018).

\bibitem{Kivelson2018} Dodaro, J. F. \&  Kivelson, S. A. Generalization of Anderson's Theorem for Disordered Superconductors, {\it Phys. Rev. B} {\bf 98}, 174503 (2018).

\bibitem{Kang2016} Kang, J. \& Fernandes, R. M.  Robustness of quantum critical pairing against disorder, {\it Phys. Rev. B} {\bf 93}, 224514 (2016).

\bibitem{Xiao1990PRB} Xiao, G., Cieplak, M. Z., Xiao, J. Q. \&  Chien, C. L.  Magnetic pair-breaking effects: Moment formation and critical doping level in superconducting La$_{1.85}$Sr$_{0.15}$Cu$_{1-x}$A$_x$O$_4$ systems (A=Fe,~Co,~Ni,~Zn,~Ga,~Al), {\it Phys. Rev. B} {\bf 42}, 8752-8755 (1990).

\bibitem{MazinSPM2008} Mazin, I. I., Singh, M. D., Johannes, D. J. \&  Du, M. H.  Unconventional Superconductivity with a Sign Reversal in the Order Parameter of LaFeAsO$_{1-x}$F$_x$,  {\it Phys. Rev. Lett. } {\bf 101}, 057003 (2008).

\bibitem{Mazin2009PhysicaC} Mazin, I. I.  \&  Schmalian, J. Pairing Symmetry and Pairing State in Ferropnictides: Theoretical Overview, {\it Physica C} {\bf 469}, 614-627 (2009).

\bibitem{Hirschfeld2011ROPP}  Hirschfeld, P. J., Korshunov, M. M. \&  Mazin, I. I. Gap
Symmetry and Structure of Fe-Based Superconductors, {\it Rep.
Prog. Phys.} {\bf 74}, 124508 (2011).

\bibitem{Chubukov2012ARCMP} Chubukov, A. Pairing Mechanism in Fe-based Superconductors, {\it Ann. Rev. Cond. Matt. Phys.} {\bf 3}, 57-92 (2012).

\bibitem{Chubukov-Hirschfeld-PT_2015} Chubukov, A.  \&  Hirschfeld,P. J. Iron-based superconductors, seven years later,  {\it Physics Today} {\bf 68}, 6, 46-52 (2015).

\bibitem{Hirschfeld2016} Hirschfeld, P. J. Using gap symmetry and structure to reveal the pairing mechanism in Fe-based superconductors, {\it Comptes Rendus Physique} {\bf 17}, 197-231 (2016).

\bibitem{Mishra2009PRB} Mishra, V. {\it et al.} Lifting of nodes by disorder in extended-s-state superconductors: Application to ferropnictides,  {\it Phys. Rev. B} {\bf 79}, 094512 (2009).

\bibitem{Onari2010}
Onari, S.  \& Kontani, H.  Nonmagnetic impurity effects and neutron scattering spectrum in iron pnictides Proceedings of the 22nd International Symposium on Superconductivity (ISS 2009), {\it Physica C} {\bf 470}, 1007-1009 (2010).

\bibitem{Onari2009PRL} Onari, S.  \&  Kontani, H.  Violation of Anderson’s Theorem for the Sign-Reversing s-Wave State of Iron-Pnictide Superconductors, {\it Phys. Rev. Lett.} {\bf 103}, 177001 (2009).

\bibitem{Kontani2010PRL} Kontani, H.  \&  Onari, S.  Orbital-Fluctuation-Mediated Superconductivity in Iron Pnictides: Analysis of the Five-Orbital Hubbard-Holstein Model, {\it Phys. Rev. Lett.} {\bf 104}, 157001
(2010).

\bibitem{Efremov2011} Efremov, D. V. {\it et al.} Disorder-induced transition between $s_\pm$ and $s_{++}$ states in two-band superconductors, {\it Phys. Rev. B} {\bf 84}, 180512
(2011).

\bibitem{FernandesChubukow2012PRB} Fernandes, R. M., Vavilov,  M. G.  \&  Chubukov, A. V.  Enhancement of $T_c$ by disorder in underdoped iron pnictide superconductors,
{\it Phys. Rev. B} {\bf 85}, 140512 (2012).

\bibitem{Xen2016} Chen, X., Mishra, V.,  Maiti, S.  \&  Hirschfeld, P. J.  Effect of nonmagnetic impurities on $s_{\pm}$ superconductivity in the presence of incipient bands, {\it Phys.
Rev. B} {\bf 94}, 054524 (2016).

\bibitem{Trevisan2018} Trevisan, T. V.,  Sch\"{u}tt, M.  \&  Fernandes, R. M.  Unconventional multi-band superconductivity in bulk SrTiO$_3$   and LaAlO$_3$/SrTiO$_3$ interfaces, {\it Phys. Rev. Lett.} {\bf 121}, 127002 (2018).

\bibitem{Kreisel2015} Kreisel, A., Mukherjee, S.  Hirschfeld, P. J.  \&
Andersen, B. M. Spin excitations in a model of FeSe with orbital ordering, {\it Phys. Rev. B} {\bf 92}, 224515 (2015).

\bibitem{BaekBuchner2015NatMat_NMR} Baek, S.-H. {\it et al.} Orbital-driven nematicity in FeSe, {\it Nat. Mater.} {\bf 14}, 210-214 (2015).


\bibitem{Chubukov2016} Chubukov, A. V., Khodas, M., \&  Fernandes, R. M.  Magnetism, superconductivity, and spontaneous orbital order in iron-based superconductors: who comes first and why? {\it Phys. Rev. X} {\bf 6}, 041045 (2016).

\bibitem{Hoyer2015} Hoyer, M.,  M. S. Scheurer, S. V. Syzranov, \&  J. Schmalian, {\it Pair breaking due to orbital magnetism in iron-based superconductors}, Phys. Rev. B {\bf 91}, 054501 (2015)

\bibitem{Golubov97} A. A. Golubov  \&   I. I. Mazin, {\it Effect of magnetic and nonmagnetic impurities on highly anisotropic superconductivity}, Phys. Rev. B {\bf 55}, 15146-15152 (1997).

\bibitem{KleinPRL2010MgB2} Klein, T. {\it et al.} First-Order Transition in the Magnetic Vortex Matter in Superconducting MgB$_2$ Tuned by Disorder, {\it Phys. Rev. Lett.} {\bf 105}, 047001 (2010).

\bibitem{Damask1963} Damask, A. C. \&  Dienes,G. J. Point Defects in Metals
(Gordon and Breach Science Publishers Ltd, London, 1963).

\bibitem{THOMPSON1969} Thompson, M. W. Defects and Radiation Damage in Metals [revised ed. 1974], Cambridge Monographs
on Physics (Cambridge University Press, Cambridge, 1969).

\bibitem{Rullier-Albenque2003PRL} Rullier-Albenque, F., Alloul, H.,  \&   Tourbot, R. Influence of Pair Breaking and Phase Fluctuations on Disordered High $T_c$ Cuprate Superconductors, Phys.
{\it Rev. Lett.} {\bf 91}, 047001 (2003).

\bibitem{VanDerBeek2013JPCS_M2S} van der Beek,C. J. {\it et al.} Electron irradiation of Co, Ni, and P-doped BaFe$_2$As$_2$ –type iron-based superconductors, {\it Journal of
Physics: Conference Series} {\bf 449}, 012023 (2013).

\bibitem{Strehlow2014}  Strehlow, C. P.  {\it et al.} Comparative study of the effects of electron irradiation and natural disorder in single crystals of SrFe$_2$(As$_{1-x}$P$_x$)$_2$ superconductor ($x=$0.35), {\it Phys. Rev. B} {\bf 90}, 020508 (2014).

\bibitem{Cho2014PRB} Cho, K.  {\it et al.} Effects of electron irradiation on resistivity and London penetration depth of Ba$_{1-x}$K$_x$Fe$_2$As$_2$ ($x \leq$0.34) iron-pnictide superconductor, {\it Phys. Rev. B} {\bf 90}, 104514 (2014).

\bibitem{Cho2016BaK} Cho, K.  {\it et al.} Energy gap evolution across the superconductivity dome in single crystals of (Ba$_{1-x}$K$_x$)Fe$_2$As$_2$,
{\it Science Adv.} {\bf 2}, e1600807 (2016).

\bibitem{Teknowijoyo2016} Teknowijoyo, S. {\it et al.} Enhancement of superconducting transition temperature by pointlike disorder and anisotropic energy gap in FeSe single crystals, {\it Phys. Rev. B} {\bf 94}, 064521 (2016).

\bibitem{SUSTreview} Cho,K,  {\it et al.} Using electron irradiation to probe iron-based superconductors,
{\it Supercond. Sci. Technol.} {\bf 31}, 064002 (2018)


\bibitem{phaseD} Tanatar, M. A.  {\it et al.} Interplane resistivity of underdoped single crystals (Ba$_{1-x}$K$_x$)Fe$_2$As$_2$ (0$\leq x\leq$0.34), {\it Phys. Rev. B} {\bf 89}, 144514 (2014).

\bibitem{Blomberg2018} Blomberg, E.~C.~ {\it et al.} Multi-band eects in in-plane resistivity anisotropy of
strain-detwinned disordered Ba(Fe$_{1-x}$Ru$_x$)$_2$As$_2$, {\it J. Phys. Cond. Matt.} {\bf  30}, 315601 (2018).

\bibitem{HaiHuWencrystals} Luo, H. Q. {\it et al.} Growth and characterization of A$_{1-x}$K$_x$Fe$_2$As$_2$ {\it (A = Ba, Sr) single crystals with $x =$0~-~0.4, {\it Supercond. Sci. Technol.} {\bf 21}, 125014 (2008).

\bibitem{SUST} Tanatar, M. A. {\it et al.} Field-dependent transport critical current in single crystals of} Ba(Fe$_{1-x}$TM$_x$)$_2$As$_2$
(TM = Co, Ni) superconductors, {\it Supercond. Sci. Technol.} {\bf 23}, 054002 (2010).

\bibitem{vdP} van der Pauw, L. J. A method of measuring specific resistivity and Hall effect of discs of arbitrary shape, {\it Philips Research Reports} {\bf 13}, 1-9 (1958).


\bibitem{yliu2014} Liu, Y. {\it et al.} Comprehensive scenario for single-crystal growth and doping dependence of resistivity and anisotropic upper critical fields in  (Ba$_{1-x}$K$_x$)Fe$_2$As$_2$ (0.22$\leq x \leq$1), {\it Phys. Rev. B} {\bf 89}, 134504 (2014)

\bibitem{Cogrowth} Ni, N. {\it et al.} Effects of Co substitution on thermodynamic and transport properties and anisotropic $H_{c2}$ in Ba(Fe$_{1-x}$Co$_x$)$_2$As$_2$ single crystals, {\it Phys. Rev. B} {\bf 78}, 214515 (2008).


\bibitem{SIRIUS} http://emir.in2p3.fr/ LSI, electron irradiation facility.


\bibitem{pseudogap} Tanatar, M. A. {\it et al.} Pseudogap and its critical point in the heavily doped Ba(Fe$_{1-x}$Co$_x$)$_2$As$_2$ from c-axis resistivity measurements,
{\it Phys. Rev. B} {\bf 82}, 134528 (2010).

\bibitem{Xigang} Luo, X. G.  {\it et al.} Quasiparticle heat transport in single-crystalline Ba$_{1-x}$K$_x$Fe$_2$As$_2$: Evidence for a $k$-dependent superconducting gap without nodes, {\it Phys. Rev. B} {\bf 80}, 140503(R) (2009).

\bibitem{Rosch1999} Rosch, A.  Interplay of Disorder and Spin Fluctuations in the Resistivity near a Quantum Critical Point, {\it Phys. Rev. Lett.} {\bf 82}, 4280-4283 (1999).

\bibitem{Stephen1977} Stephen, M. J.   \&   Grest, G. S.  Phase Transition in an Ising Model near the Percolation Threshold, {\it Phys. Rev. Lett.} {\bf 38}, 567-570 (1977).

\end{thebibliography}
\end{document}